\documentstyle[12pt,aaspp4]{article}
\input psfig.sty

\begin{document}

\title{Testing the Unified Model with an Infrared Selected Sample of
Seyfert Galaxies}

\author
  {H. R. Schmitt\altaffilmark{1,2,6,7,8,9}, R. R. J. Antonucci\altaffilmark{3},
J. S. Ulvestad\altaffilmark{1}, A. L. Kinney\altaffilmark{4,6,7},
C. J. Clarke\altaffilmark{5}, J. E. Pringle\altaffilmark{2,5}}
\altaffiltext{1}{National Radio Astronomy Observatory, P.O. Box O,
Socorro, NM 87801}
\altaffiltext{2}{Space Telescope Science Institute, 3700, San Martin Drive,
Baltimore, MD21218}
\altaffiltext{3}{University of California, Santa Barbara, Physics Department,
Santa Barbara, CA 93106}
\altaffiltext{4}{NASA Headquarters, 300 E St., Washington, DC20546}
\altaffiltext{5}{Institute of Astronomy, The Observatories, Madingley
Road, Cambridge CB3 0HA, England.}
\altaffiltext{6}{Visiting Astronomer Cerro Tololo Interamerican Observatory,
National Optical Astronomy Observatories, which is operated by AURA, Inc.
under a cooperative agreement with the National Science Foundation}
\altaffiltext{7}{Visiting Astronomer Kitt Peak National Observatory
National Optical Astronomy Observatories, which is operated by AURA, Inc.
under a cooperative agreement with the National Science Foundation}
\altaffiltext{8}{Visiting Astronomer Lick Observatory, operated by the
University of California Observatories}            
\altaffiltext{9}{Jansky Fellow}

\date{\today}

\begin{abstract}

We present a series of statistical tests done to a sample of
29 Seyfert 1 and 59 Seyfert 2 galaxies selected from mostly
isotropic properties, their far infrared fluxes and warm infrared
colors. Such selection criteria provide a profound advantage over the
criteria used by most investigators in the past, such as ultraviolet excess.
These tests were done using ground based high resolution
VLA A-configuration 3.6 cm radio and optical B and I imaging data.
From the relative number of Seyfert 1's and Seyfert 2's we calculate
that the torus half opening angle is 48$^{\circ}$.
We show that, as seen in previous papers, there is a lack of
edge-on Seyfert 1 galaxies, suggesting dust and gas along the host
galaxy disk probably play an important role in
hiding some nuclei from direct view. We find that there is no
statistically significant difference in the distribution of host
galaxy morphological types and radio luminosities
of Seyfert 1's and Seyfert 2's, suggesting that previous results showing
the opposite may have been due to selection effects.
The average extension of the radio emission of Seyfert 1's is smaller
than that of Seyfert 2's by a factor of $\sim2-3$, as predicted by the
Unified Model. A search for galaxies around our Seyferts allows us to
put a lower and an upper limit on the possible number of companions
around these galaxies of 19\% and 28\%, respectively, with no
significant difference in the number of companion galaxies between
Seyfert 1's and Seyfert 2's. We also show that there is no preference
for the radio jets to be aligned closer to the host galaxy disk axis
in late type Seyferts, unlike results claimed by previous papers.
These results, taken together, provide strong support for a Unified Model
in which type 2 Seyferts contain a torus seen more edge-on than the
torus in type 1 Seyferts.

\end{abstract}


\section {Introduction}

The Unified Scheme is based on the idea that the nuclear
engine is surrounded by a dusty molecular torus, with orientation angle
being the parameter which determines whether an AGN is perceived by
observers as a Seyfert~1 or as a Seyfert~2 (Antonucci 1993; Urry \&
Padovani 1995, Wills 1999). This scenario is supported by several
lines of evidence, like
the detection of polarized broad emission lines in Seyfert 2's
(Antonucci \& Miller 1985; Miller \& Goodrich 1990; Tran 1995; Kay
1994), or a deficit of ionizing photon in Seyfert 2 galaxies,
which indicates that the ionizing source is hidden from direct view
(Neugebauer et al. 1980; Wilson, Ward \& Haniff 1988;
Kinney et al. 1991, Storchi-Bergmann, Wilson \&
Baldwin 1992; Schmitt, Storchi-Bergmann \& Baldwin 1994).  Another strong
argument for the Unified Model is the collimated escape of radiation from
the nuclear region, detected as Narrow Line Regions
with conical shapes in Seyfert 2 galaxies and halo-like shapes in
Seyfert 1's (Pogge 1988a,b,1989; Schmitt \& Kinney 1996; Mulchaey, Wilson
\& Tsvetanov 1996a,b;
Ferruit, Wilson \& Mulchaey 2000; Haniff, Wilson \& Ward 1988; Capetti
et al. 1995, 1996), as well as linear extended, jet like radio emission
(Ulvestad \& Wilson 1984a,b, 1989; Nagar et al. 1999; Schmitt et al. 2000).

It is now accepted that the Unified Model is correct to zeroth order,
i.e. that it applies qualitatively to at least some large fraction of
the objects being studied. However, there have been some
observational results claiming intrinsic statistical
differences between Seyfert 1's and Seyfert 2's considered as populations,
which have been used as arguments against the Unified
Model. For example, Malkan, Gorjian \& Tam (1998) found that Seyfert 1's
usually reside in earlier type host galaxies compared to Seyfert 2's. They
also found that Seyfert 2's have a higher incidence of dust lanes than
Seyfert 1's. Older papers, like Meurs \& Wilson (1984) reported that in their
sample Seyfert 2's have higher radio luminosities than Seyfert 1's, and
Heckman et al. (1989) found that Seyfert 2's have higher molecular gas
masses (from CO(1-0)) than Seyfert 1's.

Historically, the main stumbling block in testing and
developing Unified Models has been the tendency to compare samples of
AGN being unified which were {\it not} preselected by some isotropic
property.  A spectacular example of this effect can be seen in the case
of samples selected by their ultraviolet excess.
According to the model, we see perhaps 1\% of the nuclear featureless
continuum by reflection in Seyfert 2's, whereas we see the
whole thing in Seyfert 1's.  Many famous Seyferts were selected by UV
excess, like the Markarian objects, and thus by this hypothesis the
Seyfert 2's were selected from two orders of magnitude higher on the
luminosity function than the Seyfert 1's. Another possibility is that
these Seyfert 2's have something else around their nuclei,
like star formation, which would increase their ultraviolet
emission.

We believe that the reason why some papers have been finding differences
in some apparently isotropic properties of Seyfert 1's and Seyfert 2's
is mostly due to the samples they use and not to intrinsic differences
between these two populations. For example, a study done by Maiolino
et al. (1997), using a larger and well defined sample than that
of Heckman et al. (1989), shows that both Seyfert types have similar
CO masses. Also, Rush, Malkan \& Edelson (1996) showed that both
types have similar radio luminosities when improved selection
criteria are used.

This paper presents a series of statistical tests done to a sample
of Seyfert galaxies selected from a mostly isotropic property, their far
infrared flux and color. We use these tests to address several problem related
to the Unified Model. The sample and data used for these tests are presented
in Section 2. Section 3 presents the comparison between the inclinations of
the host galaxies of Seyfert 1's and Seyfert 2's and their morphological types.
Section 4 shows the comparison of their radio luminosities and extensions
of the radio emission. Section 5 presents the study the percentage of
Seyferts  galaxies with companions, while Section 6 deals with the
analysis of previous suggestions that the inclination
of the torus relative to the host galaxy disk depends on the
host galaxy morphological type. A summary is given in Section 7.

\section{Sample and data}

As pointed out above, in order to be able to make a fair comparison
between Seyfert 1's and Seyfert 2's and correctly address problems
related to the Unified Model, it is necessary to use a sample selected
from an isotropic property, believed to be independent of the
orientation of the putative torus relative to the line of sight.
One of the best ways to select such a sample is based on the far
infrared properties of the galaxies.  According to the Pier \& Krolik
(1992) torus models, the circumnuclear torus radiates nearly
isotropically at 60$\mu$m, so a sample selected in this way should be
relatively free from selection effects. Any starlight energy reprocessed
to 60$\mu$m should not cause a big bias as long as that 60$\mu$m
component is optically thin and thus isotropic.

Given these facts, one of the best samples available for our purposes
is that of warm Seyfert galaxies defined by de Grijp et al. (1987,
1992). Their sample was selected based on the strength of their IRAS
25$\mu$m and 60$\mu$m fluxes, warm infrared colors
($-1.5<\alpha(25/60)<0$) to exclude starbursts as much as possible, and
high galactic latitude ($|b|>20^{\circ}$). There has been extensive
followup spectroscopy of this sample and their spectral classifications
are very secure. Also, Keel et al. (1994) showed that the [OIII] and
60$\mu$m luminosities of this sample have similar distributions.  Since
[OIII] is believed to be an isotropic quantity, because it
originates outside the torus, this result indicates that the 60$\mu$m
luminosity is an isotropic property, and consequently this sample
should be relatively unbiased.

We have to point out, however, that this sample is not complete, since any
Seyfert galaxy with infrared colors cooler than our criteria or for which
there is no good quality IRAS data, are missed. Aside from any possible
stellar contribution, any wavelength dependent far-IR anisotropy could make
an object look too cool to be selected. In fact, according to Krolik \&
Lepp (1989) the torus may be optically thick up to 50$\mu$m, which
means that the 25$\mu$m flux, used in the sample selection, is not
completely isotropic. This is not unlikely, with X-ray columns
typically N(H)$\sim10^{24}$cm$^{-2}$ or higher (Maiolino et al. 1998;
Risaliti, Maiolino \& Salvati 1999). Some encouragement here
comes from evidence of a low absorption per unit column density in AGN
(see Maiolino et al. 2001a,b).

The sample used here corresponds to all Seyfert galaxies with 
z$\leq$0.031 in the de Grijp et al. (1987, 1992) catalogs, giving a total of 88
galaxies, 29 Seyfert 1's and 59 Seyfert 2's. It was primarily selected
to study the orientation of radio jets relative to their host galaxies
disks by Kinney et al. (2000), who showed that these Seyfert 1's and
Seyfert 2's have similar 60$\mu$m luminosity distributions. This indicates
that neither of the Seyfert types is biased towards higher luminosities.

From the relative number of Seyfert 1's and Seyfert 2's in our sample
we can calculate that the torus half opening angle is $\theta=48^{\circ}$.
This is done assuming that the galaxy is recognized as a Seyfert 1 if the
observer see inside the torus, or as a Seyfert 2 otherwise. We discuss
some possible complications to this scenario in Section 3.
This value is between the ones obtained by Osterbrock \& Mart\'el (1993)
from the relative number of Seyfert 1's and 1.5's ($\theta=38^{\circ}$)
and Seyfert 1.8's and 1.9's ($\theta=56^{\circ}$) in the CfA sample.
Values similar to these can be obtain using the Palomar survey
(Ho, Filippenko \& Sargent 1997). The de Grijp et al. (1992)
classification separated the galaxies only in Seyfert 1's and Seyfert 2's.
It is possible that Seyfert 1.8's and 1.9's were divided among these two
classes, which would explain the fact that our torus half opening angle is
in the middle of the two values obtained when using samples split into
finer details.

Another important point in this study
is the use of high quality radio and broad-band optical images
obtained and measured in a homogeneous way. Data collected from the
literature can also influence the results, since it
usually is inhomogeneous, involves data of different
quality, besides the fact that different authors are likely to measure
the same thing using different techniques.
Tables 1 and 2, for Seyfert 1's and 2's, respectively, present the data
used in our analysis.

Using the galaxies radial velocities from Tables 1 and 2, and assuming
H$_0=75$ km~s$^{-1}$~Mpc$^{-1}$ (which will be used throughout this paper),
we found that Seyfert 1's and Seyfert 2's have indistinguishable
distance distributions. A KS test shows that two samples drawn from the
same parent population would differ this much 99.99\% of the time.

We also compared if Seyfert 1's and Seyfert 2's have the same space
distribution using the V/V$_{max}$ technique (Schmidt 1968), based on the
25$\mu$m and 60$\mu$m fluxes given by de Grijp et al. (1992), and the
detection limits 0.1~Jy at 25$\mu$m and 0.45~Jy at 60$\mu$m. The average
and standard deviation of V/V$_{max}$ for the 25$\mu$m data is 0.146$\pm$0.154
for Seyfert 1's and 0.111$\pm$0.144 for Seyfert 2's, while for the 60$\mu$m
data it is 0.327$\pm$0.273 and 0.251$\pm$0.214, for Seyfert 1's and Seyfert
2's, respectively. Applying a KS test to the V/V$_{max}$ distributions
obtained from the 25$\mu$m data shows that Seyfert 1's and Seyfert 2's
have similar distributions. Two samples drawn from the same parent population
would differ this much 52\% of the time. A similar result is obtained
for the V/V$_{max}$ distributions obtained from the 60$\mu$m data, with the
KS test giving a 63\% probability. These results show that both Seyfert
types have similar space distributions, and are not biased towards one type of
Seyfert being more distant than the other. One important result
from this analysis is the fact that the V/V$_{max}$ distribution have values
smaller than 0.5, which shows that the space distributions are skewed towards
nearby objects. This result is expected, since the sample includes only
galaxies with z$\leq0.031$.

The broad-band imaging data used here were presented by Schmitt \&
Kinney (2000). The host galaxy ellipticities ($e=1-b/a$) and major axis
position angles (PA$_{MA}$) were obtained by fitting ellipses to the
images of the galaxies. Further details about these measurements are
given by Kinney et al. (2000) and Schmitt \& Kinney (2000).
The Morphological Types used here were obtained mostly from
de Vaucouleurs et al. (1993) and Malkan et al. (1998). Since these
large compilations have already classified the galaxies rather homogeneously,
we gave preference to them and avoided reclassifying the galaxies.
The exceptions were the galaxies for which we could not find a previous
classification, in which case we classified them ourselves.
We also checked the galaxies smaller than 30$^{\prime\prime}$,
since older lower resolution images could have resulted in their
misclassification as earlier type systems, but none of the previous
classifications changed. 

High resolution VLA A-configuration 3.6cm radio images were available
for 75 galaxies in our sample, and PKS2048--57 had 3.6cm data from ATCA.
Data for 36 of these galaxies was obtained by us in a snapshot survey,
while for 19 other galaxies we retrieved and reduced data from the VLA
archive (Schmitt et al. 2001). Only TOL1238--364 was not
detected. Data for the remaining 20 galaxies were
obtained from Nagar et al. (1999), who reduced and measured
their data in a way similar to ours. For those galaxies with extended
radio structures, their emission was decomposed into individual components
by fitting gaussians to them. The position angle of the extended
emission (PA$_{\rm RAD}$) and the extension of the radio emission
was measured between the central position of these Gaussians.
For more details about the measurements, see Schmitt et al. (2001),
Kinney et al. (2000) and Nagar et al. (1999).

\section{Host galaxies inclinations and morphological types}

Figure 1 shows the observed distribution of the ratio between the host
galaxies minor to major axis lengths (b/a). The observed distribution has
a deficit of Seyfert 1 galaxies with b/a$<$0.5, while in Seyfert 2's
this does not seem to happen. The lack of galaxies with b/a$<$0.2 is
due to the thickness of the disk, which accounts for the fact
that even when seen edge-on a galaxy has b/a$>$0 (Hubble 1926).
Comparing the b/a distribution for
Seyfert 1's and Seyfert 2's using a Kolmogorov-Smirnov test (KS test) we find
that they are significantly different, giving the probability that
two samples drawn from the same
parent population would differ this much only 4.7\% of the time.

This result is similar to the one found by Keel (1980), who was
the first to discover a deficiency of edge-on Seyfert 1
galaxies (see also Lawrence \& Elvis 1982).
This has been confirmed by more recent results from
Maiolino \& Rieke (1995) and Simcoe et al. (1997), based on larger samples.
Also Kinney et al. (2000) showed from the modeling of the 3-dimensional
distribution of jets relative to their galaxies disk, based on the observed
distribution of host galaxy disk inclinations and $\delta$'s (the difference
between the position angles of the radio jet and the host galaxy major axis),
that there is an inconsistency in the model. This inconsistency
can only be solved if one assume that Seyfert 1's whose host galaxies
are highly inclined are seen as Seyfert 2's, regardless of the orientation
of any nuclear torus. We do not expect to see ionization cones in this case.

Although this result is in principle not expected from the Unified Model,
it does not necessarily contradict it. The papers cited above suggested that,
in the case of edge-on Seyfert galaxies, the gas and dust along the host
galaxy disk can act in the same way as a circumnuclear torus, blocking
the direct view of the Broad Line Region, thus leading to a classification
as a Seyfert 2 galaxy.

The comparison between the Morphological Types (T) of the host galaxies
of Seyfert 1's and Seyfert 2's is presented in Figure 2, which shows that
the two distributions are very similar. In fact, the KS test shows that
two samples selected from the same parent population would differ this
much 80\% of the time. This result is in contrast to the result from Malkan
et al. (1998). However, as explained in the Introduction, their result
is most likely due to selection effects, since their galaxies were
not selected following an isotropic criterion (Antonucci 1999).

\section{Radio luminosities and extensions}

Figure 3 presents the comparison between the distributions of the
radio 3.6 cm luminosities of Seyfert 1's and Seyfert 2's.
The two groups of galaxies have similar distributions,
as expected from the Unified Model. The KS test shows that two samples
selected from the same parent population would differ this much 11\%
of the time.

Figure 4 shows the distributions of the logarithm of the extension
of the 3.6cm radio emission in Seyfert 1's (top) and Seyfert 2's (bottom).
Only 9 out of 26 Seyfert 1's (35\%) show extended emission, and the rest are
unresolved, as indicated by arrows in the figure. In the case of Seyfert 2's,
28 out of 48 galaxies (58\%) have extended radio emission.

Given the fact that 50\% of the galaxies in our sample are unresolved,
we had to use survival analysis techniques to compare the two distributions.
We used the IRAF package ``statistics'' inside STSDAS, which contains
a series of statistical tests for censored data,
created based on the package ASURV (Isobe \& Feigelson 1990,
LaValley, Isobe \& Feigelson 1992).

The mean and standard deviations of the extension of the radio emission
were calculated using the Kaplan-Meier estimator, which gave
148$\pm$65 pc and 348$\pm$97 pc for Seyfert 1's and Seyfert 2's, respectively.
We should point out that the Kaplan-Meier estimator only works if the
censoring is random, meaning that the probability of a measurement being
censored does not depend on the value of the variable. In principle this
requirement seems to rule out the possibility of using this
estimator, since we cannot detect emission less extended than
$\sim0.25^{\prime\prime}$. However, since the galaxies are at
different distances, this angular size  corresponds to different physical
sizes for different galaxies, thus randomizing the censoring.

We also compared whether Seyfert 1's and Seyfert 2's have similar
distributions of the extension of the radio emission, using the
Gehan-Wilcoxon test. This test assumes that the censoring patterns are
the same in the two samples, which is our case. We obtain a probability
of 4.2\% that the two samples are drawn from the same parent population.

The results obtained from these tests confirm what a visual inspection
of Figure 4 suggests, that the extension of the radio emission in
Seyfert 1 galaxies is smaller than that of Seyfert 2's. The fact that
Seyfert 1's have smaller extended radio emission is not due to
a distance effect, since we showed that both Seyfert types have
similar distance and space distributions. We extend this analysis
by applying the V/V$_{max}$ test to the extension of the radio
emission. Excluding the galaxies with upper limits from the sample and
applying a KS test to the V/V$_{max}$ distributions of Seyfert 1's and
Seyfert 2's, we find that they do not differ significantly. Two samples
drawn from the same parent population would differ this much 72\% of the
time. In the case when we consider all galaxies, assuming those with size
upper limits have V/V$_{max}=1$, the two samples still have similar
distributions. The KS test gives an 11\% probability that two samples drawn
from the same parent population would differ this much. These results confirm
one of the predictions from the Unified Model, that, since the jets
are seen end-on in Seyfert 1 galaxies, projection effects should make
their jets look smaller than those of Seyfert 2's which are seen more edge-on.

\section{Number of companion galaxies}

Two of the major concerns in the study of AGN's are the origin of the gas
which fuels the nuclear black hole and the mechanisms that make this gas
move from galactic scales down to the inner $\sim$1 pc region of the galaxy.
In the case of spiral galaxies, gas in their disks can be a natural source
to fuel the AGN, while in ellipticals an external source seems to be
necessary. Several mechanisms have been suggested to explain how it is
possible to transport gas from the disk of a spiral
galaxy to its nucleus, like interactions (Gunn 1979; Hernquist 1989;
Hernquist \& Mihos 1995), or bars (Schwartz 1981; Norman 1987; Shlosman,
Frank \& Begelman 1990). A review about this subject is given by Combes (2001).

The influence of interactions on the fueling of AGN has been the topic
of several papers, but so far there is no consensus about this subject.
Dahari (1984) was the first one to make a large compilation of
possible companions around Seyferts,
finding that these galaxies have an excess of companions relative
to normal galaxies. The same
result was obtained by Rafanelli, Violato \& Baruffolo (1995),
Laurikainen et al. (1994) and MacKenty (1989, 1990).
On the other hand, Fuentes-Williams \& Stocke (1988), Bushouse (1986) and
De Robertis, Yee \& Hayhoe (1998) found that there is no detectable difference
in the environments of their samples of
Seyfert and normal galaxies. Yet another intriguing
result was obtained by Laurikainen \& Salo (1995) and Dultzin-Hacyan et al.
(1999), who showed that Seyfert 2's have a larger number of companions
when compared to normal galaxies, while Seyfert 1's do not.
Part of these conflicting results may be due to the way these
papers selected their samples and control samples (Heckman 1990;
Osterbrock 1993).

We used our broad band images, NED and the Digitized Sky Survey plates (DSS)
to search for companions
around the galaxies in our sample. Although we cannot do a proper
comparison between the frequency of companions in these galaxies
compared to normal ones, because we did not observe a control sample and
cannot apply the same kind of statistical analysis done by Dahari (1984),
we can at least put an upper and a lower limit on the number of galaxies
with companions in our sample for each Seyfert type.

Several parameters have been used to determine if a galaxy has a possible
companion or not. We adopt here the ones used by Rafanelli et al. (1995),
which was determined based on catalogs of interacting and pair
galaxies. A galaxy is considered a possible companion if its distance
to the galaxy of interest is smaller than 3 times the diameter of that
galaxy (3D) and the difference in brightness between them
is smaller than 3 magnitudes ($|\Delta m|\leq$ 3 mag). These criteria
alone are not enough to determine whether two galaxies actually are
companions, but they can be used to at least put an upper limit on this
number. The definitive criterion to consider two galaxies as companions,
according to Rafanelli et al. (1995) is the difference between 
the radial velocities of the two components, which has to be smaller
than $|c\Delta z|\leq$1000 km s$^{-1}$.

Following the Rafanelli et al. (1995) criteria, we searched our images
for possible companion galaxies around our Seyferts. Most of the search
was conducted on the I band images of Schmitt \& Kinney (2000), which
were deeper than their B band images. These images are sensitive enough to
ensure that all galaxies with $| \Delta I | \leq 3$ mag were detected.
The faintest galaxy in our sample has I$=$14.46 mag, while the typical
detection limit of the images is $\approx$22
mag arcsec$^{-2}$, about 7 magnitudes fainter. In the case of two
Seyfert 1 galaxies (MRK705 and MCG+08-11-011) we allowed the brightness
difference between the two galaxies to be slightly larger than 3 mag, since
they have bright nuclei, which increases their integrated magnitudes.

A small percentage of the galaxies, usually the closest ones, were large
enough that the 3D region exceeded the sides of our images. In these cases
we searched for companions on NED and DSS images. For the cases in which we
found companions on NED, they were also confirmed on the DSS images.
In these cases we had to use a $|\Delta B|\leq3$~mag brightness
criteria rather than $|\Delta I|\leq3$~mag, because there was no I band
information available for these galaxies. That should not affect our results,
since, according to H\'eraudeau, Simien \& Mamon (1996),
spiral galaxies have a mean color B--I$\sim2$, with a spread of $\sim0.5$
mag around this value. All the galaxies for which
we used the $|\Delta B|\leq3$~mag criteria have $|\Delta B|\sim$ 2 mag
or smaller.

Although the search for possible companions involved more than
one data source and was not perfectly homogeneous, all the galaxies with
$|\Delta I|\leq3$~mag or $|\Delta B|\leq3$~mag were detected. Our
images allowed to search for magnitude differences much fainter than
$|\Delta I|\leq3$~mag. The searches based on NED and DSS images also allowed
the detection of magnitude differences fainter than $|\Delta B|\leq3$~mag,
since they involved only the nearest galaxies, the brightest
ones (B$<13$ mag).

We also used NED to search for the radial velocities of the
possible companion galaxies. Some of the galaxies which satisfied the distance
and brightness criteria were excluded from the list of possible interacting
galaxies because their radial velocities differed by more than
1000 km s$^{-1}$. Nevertheless, we could not find radial velocity
information for all the galaxies.

The information about the distances and difference in radial velocities
and magnitudes between the Seyfert galaxies and their possible
companions is presented in Table 3. A total of 25 out of the 88 Seyfert
galaxies in our sample have galaxies with $|\Delta I|$ or
$|\Delta B|\leq3$~mag and closer than 3D from them, which puts
an upper limit of $<28\pm$6\% of possible interacting galaxies in this
sample (the uncertainty is given by Poisson statistics).
Of these 25 galaxies, 9 are Seyfert 1's and 16 are Seyfert 2's,
which gives an upper limit of possible companions of $<31\pm$10\% and
$<27\pm$7\% among Seyfert 1's and Seyfert 2's, respectively.
When we consider only the galaxies
which satisfy the brightness, distance and velocity criteria,
we find a total of 17 Seyferts with confirmed companions
in the sample. Since there is no
information on the radial velocities for 8 of the possible companion
galaxies, we can only assume that 17 is a lower limit, which gives
that the percentage of Seyferts with companions is
$>19\pm5$\%. Of these 17 galaxies, 7 are Seyfert 1's and 10 are Seyfert 2's,
which gives a lower limit on the number of companions of
$>24\pm9$\% and $>17\pm5$\%, respectively.

The number of confirmed companions in our sample is similar to the one
found by Rafanelli et al. (1995) for the CfA sample. We should also notice
that there is no apparent difference in the upper and lower percentage
of companion galaxies in Seyfert 1's and Seyfert 2's, contradicting the
results obtained by Laurikainen \& Salo (1995) and Dultzin-Hacyan et al.
(1999). A possible explanation for the larger percentage of Seyfert 2
galaxies with companions found by these two papers may be their samples.
Both papers selected their galaxies mostly from ultraviolet surveys.
As discussed above, this is not a good selection criterion, given the
fact that the nuclear source is hidden in Seyfert 2's and the ultraviolet
emission we see is either nuclear light reflected in our direction,
or originates from a source other than the AGN, like a Starburst
for example. If the ultraviolet excess observed in some Seyfert 2's
is in fact due to Starbursts, as several recent papers suggest
(Gonz\'alez-Delgado, Heckman \& Leitherer 2001; Storchi-Bergmann
et al. 2001; Schmitt, Storchi-Bergmann \& Cid Fernandes 1999;
Cid Fernandes, Storchi-Bergmann \& Schmitt 1998) this would in fact explain
their larger number of Seyfert 2's with companions, since galaxy
interaction can enhance star formation (Kennicutt et al. 1987;
Bushouse 1987; Telesco et al. 1993).

\section{Torus and host galaxy disk alignment as a function of morphological type}

Based on the assumption that the extended [OIII] emission observed in some
Seyfert galaxies, is a tracer of the torus axis, Wilson
\& Tsvetanov (1994) compared the orientation of the torus relative to the
host galaxy disk major axis in 11 of these objects. Their result suggested
that the torus axis could be aligned closer to the host galaxy rotation
axis in late type galaxies, since these galaxies presented [OIII] emission
extended preferentially perpendicular to the major axis, along the minor axis.
On the other hand, their results showed that the extended [OIII] emission can
have any orientation relative to the host galaxy major axis in early type
systems. According to these authors, these results suggested that the gas
feeding the nucleus could originate from the host galaxy
disk in late type galaxies, but have an external origin in early type systems.
However, as they pointed out, their results were just speculative, since
their sample was too small to draw significant conclusions.

Schmitt et al. (1997) repeated the comparison done by Wilson \& Tsvetanov
(1994), but instead of using extended [OIII] emission to trace the
torus axis, did it using extended radio emission.
Their results showed that it was not possible to distinguish the observed
distribution from a uniform one, implying that Wilson \& Tsvetanov (1994)
results were due to the small size of their sample. However, an analysis
presented by Nagar \& Wilson (1999), also based on the orientation of
radio jets, claimed the opposite, that Seyfert galaxies in late type hosts
actually have torus and accretion disk axis aligned close to the galaxy
rotation axis.

Figure 5 shows the observed distribution of $\delta$'s,
the difference between the position angles of the jets and the host
galaxies major axis, and morphological types for the sample described
in Sec.~2. This Figure is just a scatter plot, with no
particular correlation between the two quantities. A KS test shows
that the null hypothesis that the observed data is represented by a uniform
distribution, is confirmed at the 97\% significance level. In the case we
consider only those galaxies with T$\geq2$, the KS test still
gives a 76\% probability that the observed distribution can be represented
by a uniform distribution.

We believe that the Nagar \& Wilson (1999) results may be due to a selection
effect. They found that there is a tendency for Seyferts in types later than
Sab (T$\geq$2) to avoid close alignment of the jets with the minor axis.
However, their primary sample was composed only of Seyferts with hosts
earlier than Sab (T$\leq$2), to which they added some later type
galaxies obtained from the literature. In this way,
they ended up having a very good representation of galaxies with T$\leq2$,
but lacked galaxies with later types, which most likely influenced their
results.
      
In order to increase the number of galaxies in this comparison, we repeated
this test including all Seyfert galaxies with extended radio emission presented
by Kinney et al. (2000), which were not in our sample described in Sec.~2.
The $\delta$ versus T distribution is presented in Figure 6,
which shows the same behavior as Figure 5. A KS test also gives similar
results, showing that the null hypothesis that the observed data
can be represented by a uniform distribution is confirmed at the
84\% level, or 23\% when we consider only those
galaxies with T$\geq2$. These results are expected, since Clarke et al.
(1998), Nagar \& Wilson (1999) and Kinney et al. (2000) showed that
there is no correlation between the orientation of the jet axis
and host galaxy disk. The possible implications of this result for
the structure of the accretion disk and feeding of the black hole
are discussed by Kinney et al. (2000).

\section{Summary}

In this paper, we used high resolution radio 3.6cm and
optical B and I images of a sample of Seyfert galaxies
selected from their far infrared properties, which are believed 
to be relatively isotropic properties of these galaxies, since they are
relatively unbiased. Our sample
selection makes it one of the best ones to study the
Unified Model, since it is unbiased towards the orientation
of the accretion torus. 

From the relative number of Seyfert 1's and Seyfert 2's we calculate that
the torus half opening angle is $\theta=48^{\circ}$. This value is
similar to the one found by Osterbrock \& Mart\'el (1993) for the CfA
sample.

Our results showed that there is a deficiency of Seyfert 1's
in edge-on galaxies. The comparison between the distribution of 
the ratio between the host galaxy minor to major axis lengths
shows that Seyfert 1's and Seyfert 2's are statistically different,
with the Seyfert 2's having more edge-on galaxies. This result is in
line with previous results, which suggest that dust and gas along the
host galaxy disk can play an important role in hiding the nucleus
from direct view in edge-on Seyferts.

The comparison of the morphological types of the host galaxies
does not show a statistically significant difference between Seyfert 1's
and Seyfert 2's. Previous results showing the opposite may
have been due to selection effects. We also find that there is no significant
difference in the radio 3.6cm luminosities of these two groups.

Using survival analysis techniques, we showed that the extension of
the radio emission in Seyfert 1's is smaller than that of Seyfert 2's
by a factor of 2--3.
This result is in accordance with the Unified Model, which predicts
that the jets are seen end-on in Seyfert 1 galaxies and therefore should
be shortened by projection.

We used our images to search for possible companions around our galaxies,
following the criteria described by Rafanelli et al. (1995). Although
we could not compare the number we obtained with that of normal galaxies,
because we did not observe a control sample, we were able to determine
limits to the percentage of galaxies with companions in our sample
and compare their frequency in Seyfert 1's and Seyfert 2's. We found
that between 19\% and 28\% of the galaxies in our sample have companions,
with no significant difference between Seyfert 1's and Seyfert 2's.
Previous results showing that there is a higher percentage of Seyfert 2's
with companions than Seyfert 1's, may have been due to the samples
used to perform those studies.

We also showed that there is no preference for the torus
axis of Seyferts in late type galaxies to be aligned preferentially
close to their host galaxy rotation axis (minor axis) as shown by previous
works. The observed distribution suggests that there is no correlation
between the jet and the disk direction, as shown by Kinney et al. (2000).

\acknowledgements 

We would like to acknowledge the hospitality and help from the staff at
CTIO, KPNO and Lick Observatories during the observations. We also would
like to thank the anonymous referee for
usefull comments. This work was supported by NASA grant AR-8383.01-97A.
This research made use of the NASA/IPAC Extragalactic Database (NED),
which is operated by the Jet Propulsion Laboratory, Caltech, under
contract with NASA. We also used the Digitized Sky Survey, which was
produced at the Space Telescope Science Institute under U.S. Government
grant NAGW-2166. The National Radio Astronomy Observatory is a facility
of the National Science Foundation operated under cooperative agreement
by Associated Universities, Inc.

\clearpage
\newpage

\begin{figure}
\psfig{figure=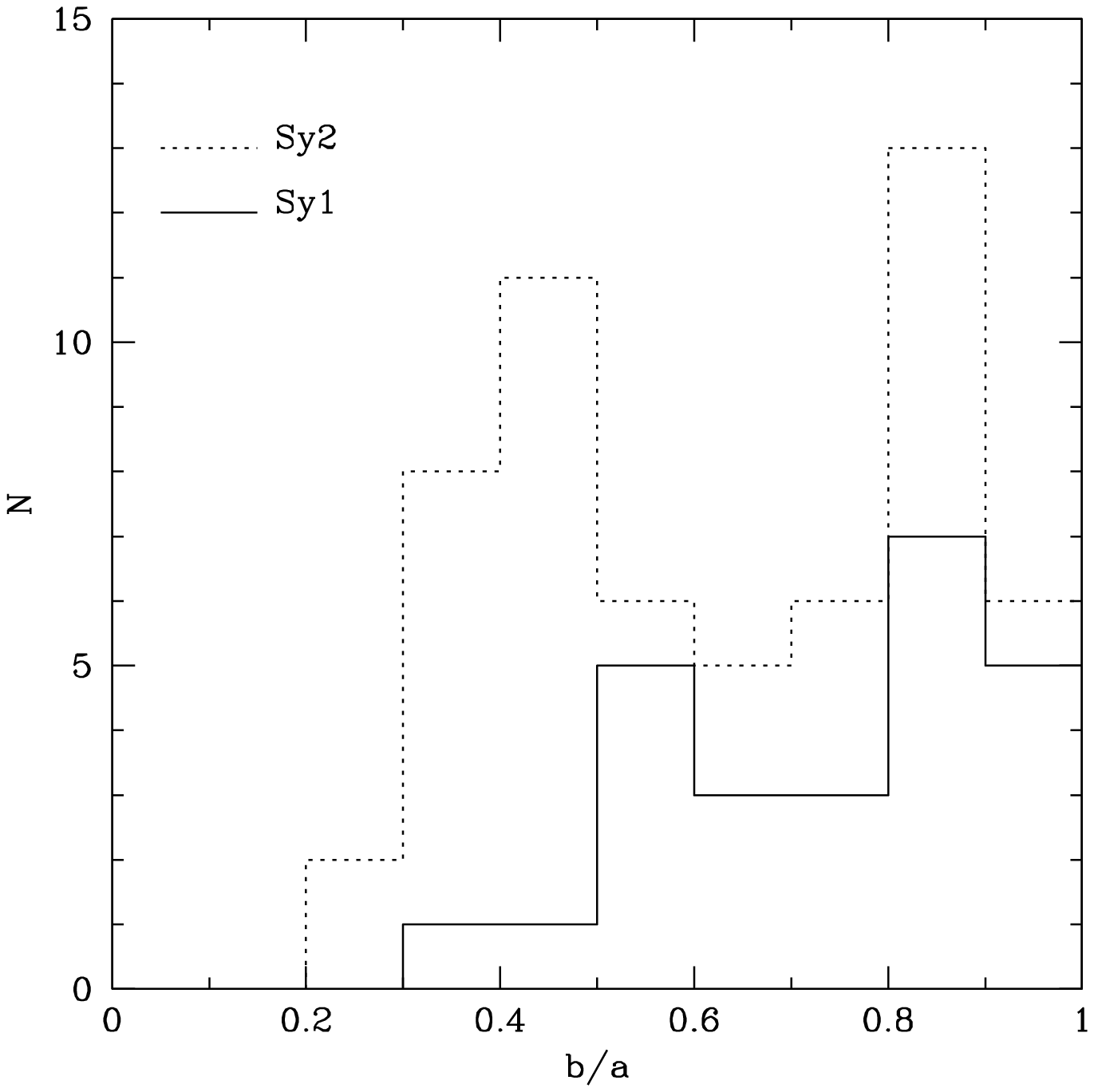,width=20cm,height=20cm}
\caption{Histogram of the ratio between the semi-minor and semi-major
axes of the host galaxies of
Seyfert 1's (solid line) and Seyfert 2's (dotted line).}
\end{figure}

\begin{figure}
\psfig{figure=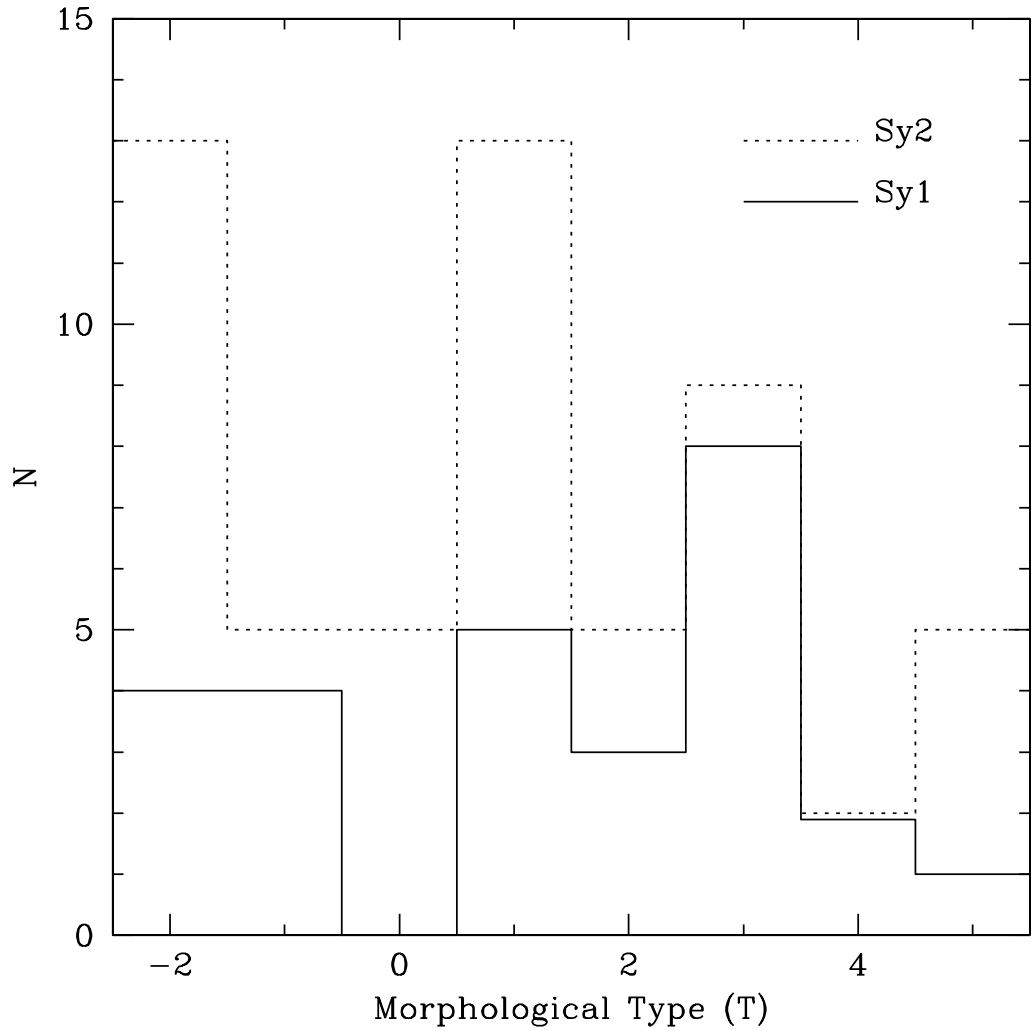,width=20cm,height=20cm}
\caption{Histogram of the Morphological Types of Seyfert 1's (solid line)
and Seyfert 2's (dotted line). T=-2 corresponds to morphological type S0,
T=0  S0/a, T=2  Sab, T=4  Sbc. One of the Seyfert 2 galaxies in our sample
(IRAS\,01475-0740) is classified as an elliptical, T=-5, but we include
it in the T=-2 bin.}
\end{figure}

\begin{figure}
\psfig{figure=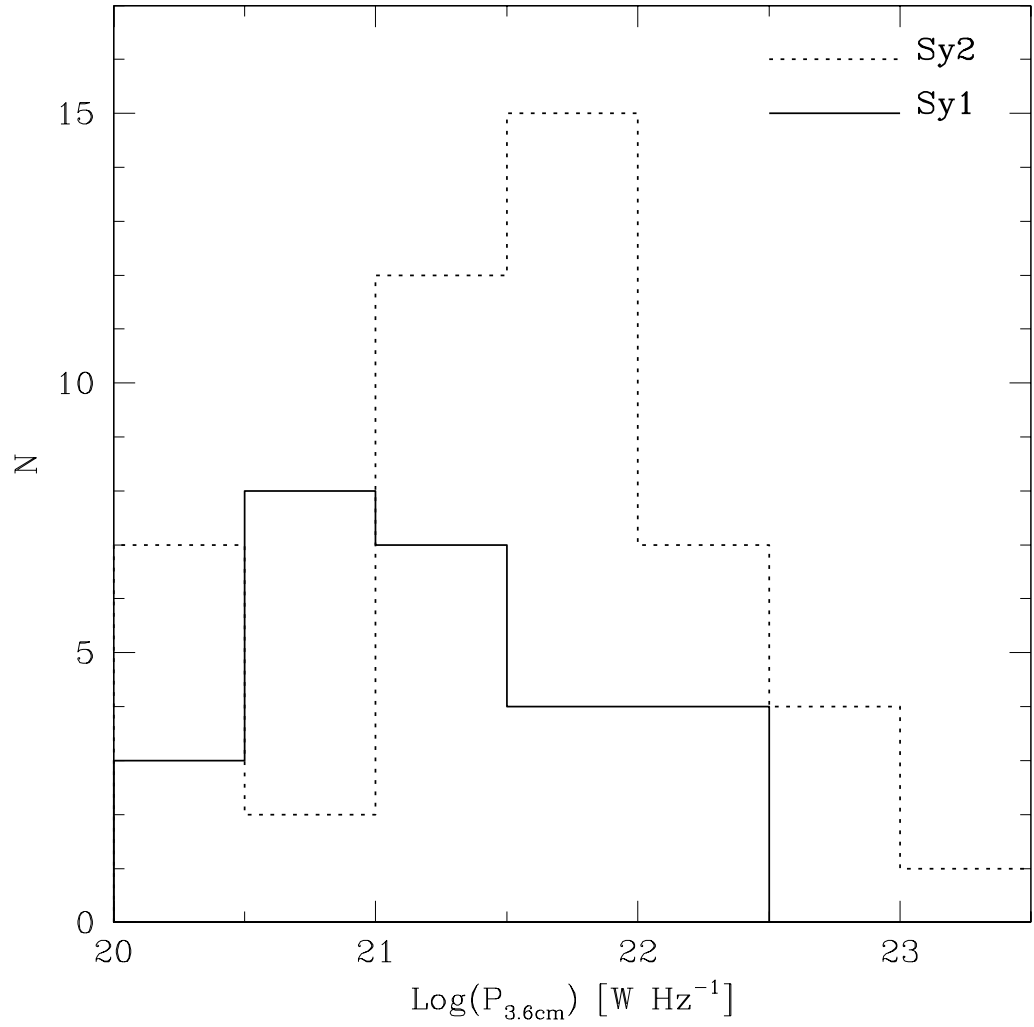,width=20cm,height=20cm}
\caption{Histogram of the logarithm of the 3.6cm radio luminosities
of Seyfert 1's (solid line) and Seyfert 2's (dotted line).}
\end{figure}

\begin{figure}
\psfig{figure=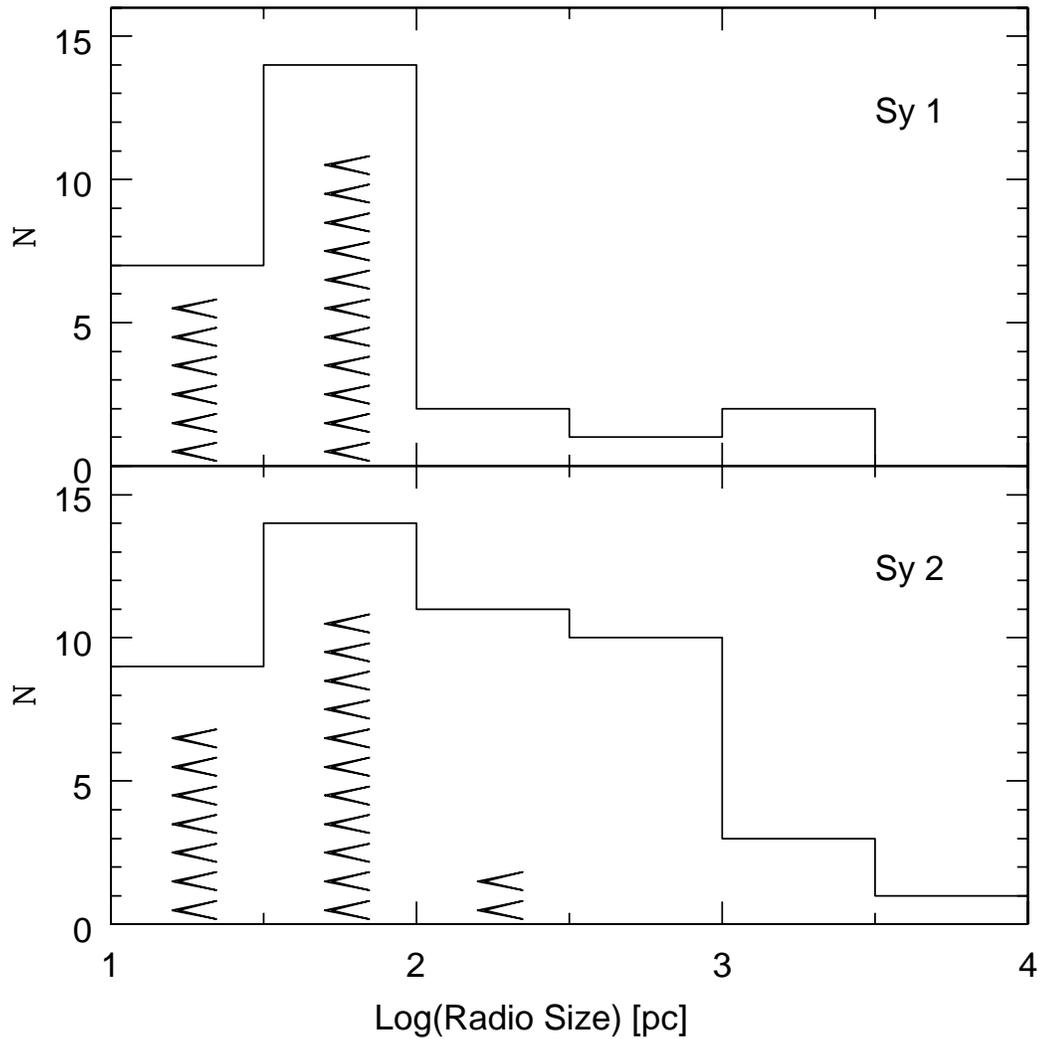,width=20cm,height=20cm}
\caption{Comparison between the distributions of the Logarithm of the
extension of the radio emission in Seyfert 1's (top) with that of
Seyfert 2's (bottom). The histograms represent the total number of
galaxies inside each bin, adding up those with detected extended
emission and upper limits. The arrows represent the number of
galaxies with upper limits inside each one of the bins.}
\end{figure}

\begin{figure}
\psfig{figure=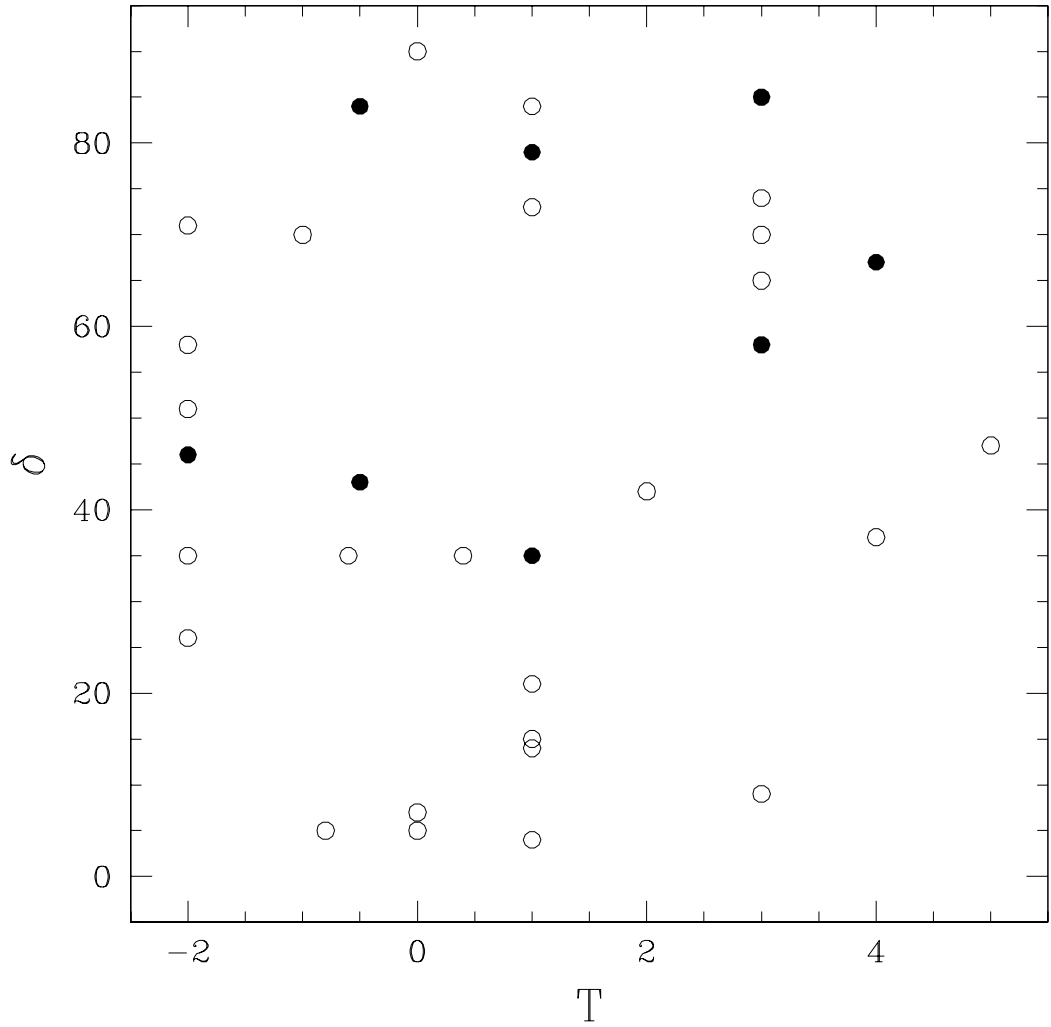,width=20cm,height=20cm}
\caption{Distribution of $\delta$'s, the difference between the position angle
of the extended radio emission and the position angle of the host galaxy major
axis, as a function of morphological types (T).
Seyfert 1's are represented by solid dots and Seyfert 2's by open dots.}
\end{figure}

\begin{figure}
\psfig{figure=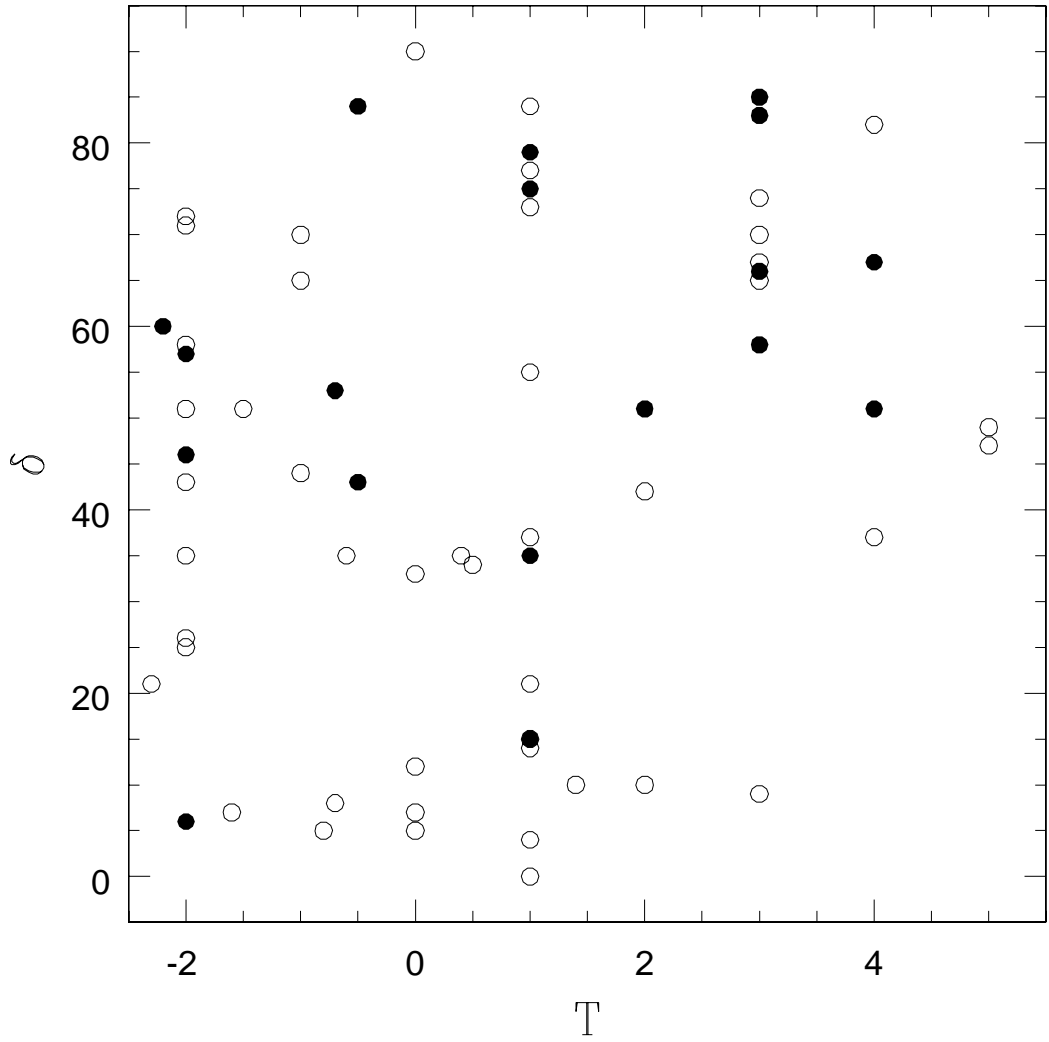,width=20cm,height=20cm}
\caption{Distribution of $\delta$'s as a function of morphological type.
This plot includes all galaxies from Figure 5, plus all the Seyfert
galaxies from Nagar \& Wilson (1999) and Kinney et al. (2000)
known to have extended radio emission. Symbols as in Figure 5.}
\end{figure}

\clearpage
\newpage
\begin{tiny}
 
\begin{deluxetable}{rlrrrrrrr}
\tablewidth{0pc}
\tablecaption{Seyfert 1 Galaxies Properties}
\tablehead{\colhead{Number}&\colhead{Name}&\colhead{Velocity}&\colhead{b/a}&
\colhead{Radio Extent}&\colhead{Log(L$_{3.6cm}$)}&\colhead{T}&
\colhead{$\delta$}&\colhead{Reference}\cr
& &(km s$^{-1}$) & &(pc)&(W~Hz$^{-1}$)& &(deg)&}
\startdata
27&MRK\,359          	&5012& 0.83& 170  & 20.43  & 4    & 67  & 2  \cr
47&MRK\,1040 	   	&4927& --- & $<$32& 20.76  & 4    & --  & 3  \cr
68&UGC\,2514         	&3957& 0.57& 69   & 20.30  & -0.5 & 84  & 3  \cr
174&MCG\,-05-13-017  	&3790& 0.81& $<$24& 20.88  & -0.3 & --  & 3  \cr
203&UGC\,3478        	&3828& 0.38& $<$25& 20.64  & 3    & --  & 3  \cr
209&MRK\,6           	&5537& 0.59& 437  & 22.29  & -0.5 & 43  & 3  \cr
213&FAIRALL\,265     	&8844& 0.75& ---  &  ---   & 2    & --  & 1  \cr
225&MRK\,79          	&6652& 0.81& 1255 & 21.44  & 3    & 58  & 3  \cr	
227&MRK\,10	   	&8785& 0.51& $<$57& 20.69  & 3    & --  & 3  \cr
233&UGC\,4155	   	&7645& 0.77& $<$49& 21.48  & 1    & --  & 3  \cr
260&MRK\,1239 	   	&5974& 0.81& $<$53& 21.78  & -2   & --  & 4  \cr
278&NGC\,3516        	&2649& 0.81& 18   & 20.79  & -2   & 46  & 3  \cr
286&NGC\,3783	   	&2550& 0.93& $<$17& 21.03  & 1.5  & --  & 3  \cr
292&MRK\,766         	&3876& 0.85& 67   & 21.44  & 1    & 35  & 3  \cr
301&NGC\,4593        	&2698& 0.71& $<$17& 20.51  & 3    & --  & 3  \cr
309&MCG\,-02-33-034  	&4386& --- & $<$28& 20.69  & --   & --  & 1  \cr
324&MCG\,-6-30-15    	&2323& 0.65& $<$38& 20.01  & -2   & --  & 2  \cr
344&NGC\,5548        	&5149& --- & 44   & 21.24  & --   & --  & 3  \cr
369&UGC\,9826 	   	&8754& 0.67& $<$57& 20.51  & 5    & --  & 3  \cr
473&FAIRALL\,51	   	&4255& 0.55& ---  &  ---   & 2.8  & --  & 3  \cr
497&ESO\,143-G09	&4462& 0.67& ---  &  ---   & 2.6  & --  & 3  \cr
530&NGC\,7213	   	&1792& 0.97& $<$12& 22.11  & 1    & --  & 3  \cr
537&MRK\,915	   	&7230& 0.49& $<$47& 22.21  & 3    & --  & 4  \cr
538&UGC\,12138	   	&7375& 0.92& $<$81& 21.34  & 1    & --  & 3  \cr
540&AKN\,564         	&7195& 0.80& 316  & 21.89  & 1    & 79  & 1  \cr
590&MRK\,590	   	&7910& 0.91& $<$51& 21.63  & 1.3  & --  & 3  \cr
615&MCG\,+8-11-11    	&6141& 0.90& 1230 & 22.29  & 3    & 85  & 1  \cr
627&MRK\,705	   	&8658& 0.77& $<$56& 21.50  & -2   & --  & 3  \cr
703&UGC\,10683B      	&9234& --- & $<$60& 21.04  & -1   & --  & 3  \cr
\tablenotetext{}{Column 1: galaxy entry number in the de Grijp et al. (1987)
catalog;}
\tablenotetext{}{Column 2: galaxy name;}
\tablenotetext{}{Column 3: radial velocity;}
\tablenotetext{}{Column 4: ratio between the semi minor and semi major host
galaxy axis length;}
\tablenotetext{}{Column 5: linear extent of the radio emission;}
\tablenotetext{}{Column 6: logarithm of the 3.6cm radio luminosity (from
Schmitt et al. 2001 or Kinney et al. 2000);}
\tablenotetext{}{Column 7: host galaxy morphological type, T$<-2$ corresponds
to S0 and earlier, and T$>5$ to Sc and later;}
\tablenotetext{}{Column 8: $\delta$, the difference between the position angles
of the radio jet and host galaxy major axis (PA$_{RAD}-$PA$_{MA}$);}
\tablenotetext{}{Column 9: References from which the morphological types
were obtained. 1) estimated from Schmitt \& Kinney (2000); 2) Malkan et al.
(1998); 3) de Vaucouleurs et al. (1991); 4) NED.}
\enddata
\end{deluxetable}
 
\clearpage
\newpage

\begin{deluxetable}{rlrrrrrrr}
\tablewidth{0pc}
\tablecaption{Seyfert 2 Galaxies Properties}
\tablehead{\colhead{Number}&\colhead{Name}&\colhead{Velocity}&\colhead{b/a}&
\colhead{Radio Extent}&\colhead{Log(L$_{3.6cm}$)}&\colhead{T}&
\colhead{$\delta$}&\colhead{Reference}\cr
& &(km s$^{-1}$) & &(pc)&(W~Hz$^{-1}$)& &(deg)&}
\startdata
16&MRK\,348          	&4540& 0.94& 33   & 23.18  & 0    &  5  & 3  \cr
24&TOL\,0109-38      	&3496& 0.45& 330  & 21.53  & 0.4  & 35  & 3  \cr	
26&MRK\,1  	   	&4780& 0.64& $<$31& 21.73  & 5    & --  & 2  \cr
33&MRK\,573          	&5174& 0.86& 1003 & 21.01  & -1   & 70  & 3  \cr	
37&IRAS\,01475-0740  	&5306& 0.81& $<$34& 22.91  & -5   & --  & 2  \cr
41&ESO\,153-G20	   	&5917& 0.70& ---  &  ---   & 2.6  & --  & 3  \cr
52&ESO\,355-G25 	&5039& 0.91& ---  &  ---   & 2.4  & --  & 3  \cr
53&UGC\,2024 	   	&6714& 0.63& $<$43& 20.83  & 2    & --  & 3  \cr
57&NGC\,1068         	&1136& 0.89& 745  & 22.32  & 3    & 65  & 3  \cr	
67&MCG\,-02-08-039   	&8874& 0.57& $<$57& 21.53  & 1    & --  & 3  \cr
75&IRAS\,03106-0254  	&8154& 0.32& 854  & 22.18  & -2   & 51  & 4  \cr
78&IRAS\,03125+0119  	&7200& 0.56& $<$47& 21.99  & -2   & --  & 1  \cr
83&MRK\,607 	   	&2716& 0.38& $<$20& 20.31  & 1    & --  & 3  \cr
85&ESO\,116-G18 	&5546& 0.43& ---  &  ---   & -0.6 & --  & 3  \cr
141&IRAS\,04385-0828 	&4527& 0.48& $<$29& 21.52  & -2   & --  & 4  \cr
154&IRAS\,04502-0254 	&4737& 0.51& 107  & 20.28  & 0    &  7  & 4  \cr
156&IRAS\,04507+0358 	&8811& 0.81& 205  & 21.12  & 1    & 73  & 4  \cr
157&ESO\,33-G02      	&5426& 0.91& ---  &  ---   & -2   & --  & 2  \cr
196&MRK\,3           	&4050& 0.84& 375  & 22.44  & -2   & 58  & 3  \cr
236&MRK\,622         	&6964& 0.90& 110  & 21.24  & -2   & 35  & 3  \cr	
244&ESO\,18-G09	   	&5341& 0.86& ---  &  ---   & 5    & --  & 4  \cr
253&MCG\,-01-24-012  	&5892& 0.59& 133  & 21.82  & 5    & 47  & 4  \cr
272&NGC\,3393        	&4107& 0.87& 683  & 21.59  & 1    & 15  & 3  \cr
281&IRAS\,11215-2806 	&4047& 0.34& 403  & 21.58  & -2   & 71  & 2  \cr
282&MCG\,-05-27-013  	&7162& 0.36& 1530 & 21.71  & 1    & 84  & 3  \cr
283&MRK\,176         	&8346& --- & 135  & 21.97  & 1    & --  & 3  \cr
293&NGC\,4388        	&2524& 0.34& 2940 & 21.11  & 3    & 70  & 3  \cr
299&NGC\,4507	   	&3538& 0.84& ---  &  ---   & 3    & --  & 3  \cr
302&TOL\,1238-364    	&3285& 0.90& ---  &  ---   & 4    & --  & 3  \cr
306&NGC\,4704        	&8134& 0.92& $<$53& 21.05  & 3.5  & --  & 3  \cr
310&ESO\,323-G32     	&4796& 0.90& ---  &  ---   & -1.3 & --  & 3  \cr
313&MCG\,-04-31-030  	&2957& 0.47& 478  & 21.08  & -2   & 26  & 3  \cr
314&IRAS\,13059-2407 	&4175& 0.33& $<$27& 22.67  & 5    & --  & 2  \cr
317&MCG\,-03-34-064  	&5152& 0.68& 278  & 22.49  & 3    &  9  & 2  \cr
322&ESO\,383-G18     	&3837& 0.45& 107  & 20.73  & 0    & 90  & 4  \cr
329&NGC\,5347	   	&2335& 0.79& $<$15& 20.27  & 2    & --  & 3  \cr
340&IRAS\,14082+1347 	&4836& 0.68& 53   & 21.20  & 1    &  4  & 1  \cr
341&NGC\,5506        	&1753& 0.25& 302  & 21.79  & 1    & 21  & 3  \cr
349&IRAS\,14317-3237 	&7615& 0.76& 286  & 21.37  & 1    & 14  & 1  \cr
354&IRAS\,14434+2714 	&8814& 0.88&$<$120& 22.16  & 1    & --  & 2  \cr
377&UGC\,9944        	&7354& 0.33& 3430 & 21.83  & 3    & 74  & 4  \cr
383&IRAS\,15480-0344 	&9084& 0.89& $<$59& 22.29  & -2   & --  & 2  \cr
409&IRAS\,16288+3929 	&9091& 0.43& $<$59& 21.93  & 0    & --  & 4  \cr
418&IRAS\,16382-0613 	&8317& 0.75& $<$54& 21.49  & ---  & --  & 1  \cr
445&UGC\,10889	   	&8424& 0.42& $<$54& 21.22  & 3    & --  & 3  \cr
447&MCG\,+03-45-003  	&7292& 0.57& $<$47& 20.53  & 5    & --  & 1  \cr
471&FAIRALL\,49	   	&6065& 0.89& ---  &  ---   & 1    & --  & 2  \cr
501&FAIRALL\,341	&4887& 0.77& ---  &  ---   & -2   & --  & 3  \cr
510&UGC\,11630	   	&3657& 0.47& $<$24& 20.06  & -1   & --  & 3  \cr
512&PKS\,2048-57     	&3402& 0.84& 924  & 22.75  & -0.8 &  5  & 3  \cr
549&UGC\,12348	   	&7585& 0.29&$<$118& 21.13  & 1    & --  & 3  \cr
555&NGC\,7674        	&8713& 0.73& 422  & 22.81  & 4    & 37  & 3  \cr
594&MRK\,1058	   	&5138& 0.58& $<$33& 20.05  & 3    & --  & 2  \cr
602&NGC\,1386        	&868& 0.37& 20   & 20.24  & -0.6 & 35  & 3  \cr
634&NGC\,3281	   	&3200& 0.41& 60   & 21.57  & 2.1  & --  & 3  \cr
638&UGC\,6100	   	&8778& 0.66& $<$57& 21.12  & 1    & --  & 3  \cr
665&NGC\,4941        	&1108& 0.48& 15   & 20.02  & 2    & 42  & 3  \cr
708&ESO\,103-G35	&3983& 0.40& ---  &  ---   & -2   & --  & 4  \cr
721&NGC\,7212        	&7984& --- & 361  & 22.39  & --   & --  & 1  \cr
\tablenotetext{}{Column 1: galaxy entry number in the de Grijp et al. (1987)
catalog;}
\tablenotetext{}{Column 2: galaxy name;}
\tablenotetext{}{Column 3: radial velocity;}
\tablenotetext{}{Column 4: ratio between the semi minor and semi major host
galaxy axis length;}
\tablenotetext{}{Column 5: linear extent of the radio emission;}
\tablenotetext{}{Column 6: logarithm of the 3.6cm radio luminosity (from
Schmitt et al. 2001 or Kinney et al. 200));}
\tablenotetext{}{Column 7: host galaxy morphological type, T$<-2$ corresponds
to S0 and earlier, and T$>5$ to Sc and later;}
\tablenotetext{}{Column 8: $\delta$, the difference between the position angles
of the radio jet and host galaxy major axis (PA$_{RAD}-$PA$_{MA}$);}
\tablenotetext{}{Column 9: References from which the morphological types
were obtained. 1) estimated from Schmitt \& Kinney (2000); 2) Malkan et al.
(1998); 3) de Vaucouleurs et al. (1991); 4) NED.}
\enddata
\end{deluxetable}
 
\clearpage
\newpage

\begin{deluxetable}{llrrrrl}
\tablewidth{0pc}
\tablecaption{Galaxies with companions}
\tablehead{\colhead{Name}  & \colhead{Type} & \colhead{$|\Delta V|$}&
\colhead{$|\Delta I|$} &  \colhead{$|\Delta B|$}& \colhead{Distance}&    
\colhead{Comments}\cr
&&(km s$^{-1}$)&(mag)&(mag)&&}
\startdata
MCG\,-02-33-034  & 1 & --- &   --- & ---    & ---    & Interacting Galaxy\cr
MCG\,+08-11-011  & 1 & 180$^a$ &3.31$^e$& ---    & 65$^{\prime\prime}$W   &\cr
MRK\,705&1&---&3.05$^e$&---&31$^{\prime\prime}$S&Ring galaxy, Cartwheel like\cr
MRK\,915         & 1 & 71  &   1.55& ---    & 119$^{\prime\prime}$E  &      \cr
MRK\,1040        & 1 & 143$^b$ &   2.97& ---    & 17$^{\prime\prime}$N   &\cr
MRK\,1239        & 1 & 60$^a$  &   1.67& ---    & 71$^{\prime\prime}$S   &\cr
NGC\,4593        & 1 & 320 &    ---& 2.57$^c$   & 228$^{\prime\prime}$E  &\cr 
UGC\,2514        & 1 & --- &   2.25& ---    & 155$^{\prime\prime}$NW &      \cr
UGC\,10683\,B    & 1 & 360 &    ---& 0.60$^f$& 67$^{\prime\prime}$NW  &      \cr
FAIRALL\,341     & 2 & --- &   2.26& ---    & 165$^{\prime\prime}$W  &      \cr
IRAS\,13059-2407 & 2 & --- &   2.27& ---    & 55$^{\prime\prime}$SE  &      \cr
IRAS\,14082+1347 & 2 & --- &   2.76& ---    & 106$^{\prime\prime}$SW &      \cr
MCG\,-01-24-012  & 2 & 56  &   0.04& ---    & 79$^{\prime\prime}$N   &      \cr
MCG\,+03-45-003  & 2 & 100 &   1.97& ---    & 35$^{\prime\prime}$NW  &      \cr
MRK\,1           & 2 & 100 &    ---& 0.25$^c$   & 111$^{\prime\prime}$SE &\cr
MRK\,176         & 2 & 236 &   0.80& ---    & 33$^{\prime\prime}$W   &      \cr
MRK\,348         & 2 & 540$^d$ &   1.71& ---    & 70$^{\prime\prime}$E   &\cr
MRK\,607         & 2 & 18  &   0.27& ---    & 92$^{\prime\prime}$N   &      \cr
MRK\,1058        & 2 & --- &   2.93& ---    & 61$^{\prime\prime}$W   &      \cr
NGC\,3281        & 2 & --- &    ---& 2.38$^c$   & 500$^{\prime\prime}$S  &\cr
NGC\,3393        & 2 & --- &    ---& 2.20$^c$   & 252$^{\prime\prime}$NW &\cr
NGC\,5506        & 2 & 99  &   0.24& ---    & 227$^{\prime\prime}$NE &      \cr
NGC\,7212        & 2 & 74  &    ---& ---    & 18$^{\prime\prime}$N   & Interacting Galaxy\cr
NGC\,7674        & 2 & 139 &   1.57& ---    & 32$^{\prime\prime}$NE  &      \cr
TOL\,1238-364    & 2 & 235 &   0.16& ---    & 105$^{\prime\prime}$NE &      \cr
\tablenotetext{}{Column 1: galaxy name;}
\tablenotetext{}{Column 2: Seyfert activity type;}
\tablenotetext{}{Column 3: modulus of the difference between the radial
velocities of the two galaxies;}
\tablenotetext{}{Column 4: modulus of the difference between their
I magnitudes;}
\tablenotetext{}{Column 5: modulus of the difference between their
B magnitudes;}
\tablenotetext{}{Column 6: Distance and direction of the companion galaxy
in relation to the galaxy indicated in Column 1;}
\tablenotetext{}{Column 7: Comments.}
\tablenotetext{a}{Velocity information from Kell (1996).}
\tablenotetext{b}{Velocity information from Ward \& Wilson (1976).}
\tablenotetext{c}{B magnitude information from NED.}
\tablenotetext{d}{Velocity information from Dahari (1985).}
\tablenotetext{e}{$|\Delta I|<3$ alleviated because the galaxy is a Seyfert 1
with strong nuclear emission.}
\tablenotetext{f}{$| \Delta B |$ obtained from Schmitt \& Kinney (2000) images.}
\enddata
\end{deluxetable}
\end{tiny}

\end{document}